\documentclass[sigconf]{acmart}

\def \urlrefarch {\url{http://bit.ly/2TIMmDh}}
\def \urldataset {\url{http://bit.ly/2VQrAUU}}

\usepackage{import}
\usepackage{natbib}
\usepackage{blindtext}  %
\usepackage{enumitem}   %
\usepackage{hyperref}   %
\usepackage{cleveref}   %
\usepackage{caption}
\makeatletter
\let\MYcaption\@makecaption
\makeatother

\usepackage[font=footnotesize]{subcaption}

\makeatletter
\let\@makecaption\MYcaption
\makeatother

\usepackage[multiple]{footmisc} %

\newcommand{\footnoteurl}[2]{\footnote{{#1} last accessed #2.}} %
\usepackage{graphicx}
\graphicspath{{images/}{../images/}}
\usepackage{listings}
\usepackage{textcomp}
\usepackage{color}
\definecolor{maroon}{rgb}{0.5,0,0}
\definecolor{darkgreen}{rgb}{0,0.5,0}

\newcommand{\beginlstdelim}[3]
{%
  \def\endlstdelim{#2\egroup}%
  \ttfamily#1\bgroup\color{#3}\aftergroup\endlstdelim%
}

\lstdefinestyle{xml_style}
{
  morestring=[b]",
  moredelim=*[s][\bfseries\color{maroon}]{<}{>},
  moredelim=*[s][\bfseries\color{maroon}]{</}{>},
  morecomment=[s]{<?}{?>},
  morecomment=[s]{<!--}{-->},
}

\lstdefinestyle{json_style}
{
  morestring=[b]",
  comment=[l]{//},
  keywordstyle=\bfseries\color{maroon}
}

\lstdefinelanguage{JSON}
{
  style=json_style
}

\lstdefinelanguage{XML}
{
  style=xml_style
}
\usepackage{booktabs}
\usepackage{multirow}
\usepackage{makecell}
\newcommand{\tablefit}[1]{\resizebox{\linewidth}{!}{#1}}

\usepackage{xcolor}

  \usepackage{balance}
\title[Beware the evolving `intelligent' web service!\\An integration architecture tactic to guard AI-first components]{Beware the evolving `intelligent' web service!\\An integration architecture tactic to guard AI-first components}

\author{Alex Cummaudo}
\email{ca@deakin.edu.au}
\orcid{0000-0001-7878-6283}
\affiliation{%
\department{Applied Artificial Intelligence Inst.}
\institution{Deakin University}
\city{Geelong}
\country{Australia}
}

\author{Scott Barnett}
\email{scott.barnett@deakin.edu.au}
\orcid{0000-0002-3187-4937}
\affiliation{%
\department{Applied Artificial Intelligence Inst.}
\institution{Deakin University}
\city{Geelong}
\country{Australia}
}

\author{Rajesh Vasa}
\email{rajesh.vasa@deakin.edu.au}
\orcid{0000-0003-4805-1467}
\affiliation{%
\department{AApplied Artificial Intelligence Inst.}
\institution{Deakin University}
\city{Geelong}
\country{Australia}
}

\author{John Grundy}
\email{john.grundy@monash.edu}
\orcid{0000-0003-4928-7076}
\affiliation{%
\department{Faculty of Information Technology}
\institution{Monash University}
\city{Clayton}
\country{Australia}
}

\author{Mohamed Abdelrazek}
\email{mohamed.abdelrazek@deakin.edu.au}
\orcid{0000-0003-3812-9785}
\affiliation{%
\department{School of Information Technology}
\institution{Deakin University}
\city{Geelong}
\country{Australia}
}

\begin{abstract}
Intelligent services provide the power of AI to developers via simple RESTful API endpoints, abstracting away many complexities of machine learning.
However, most of these intelligent services---such as computer vision---continually learn with time.
When the internals within the abstracted `black box' become hidden and evolve, pitfalls emerge in the robustness of applications that depend on these evolving services. 
Without adapting the way developers plan and construct projects reliant on intelligent services, significant gaps and risks result in both project planning and development.
Therefore, how can software engineers best mitigate software evolution risk moving forward, thereby ensuring that their own applications maintain quality?
Our proposal is an architectural tactic designed to improve intelligent service-dependent software robustness. The tactic involves creating an application-specific benchmark dataset baselined against an intelligent service, enabling evolutionary behaviour changes to be mitigated. 
A technical evaluation of our implementation of this architecture demonstrates how the tactic can identify 1,054 cases of substantial confidence evolution and 2,461 cases of substantial changes to response label sets using a dataset consisting of 331 images that evolve when sent to a service.
\end{abstract}



\renewcommand{\shortauthors}{Alex Cummaudo et al.}

\acmConference[ESEC/FSE 2020]{The 28th ACM Joint European Software Engineering Conference and Symposium on the Foundations of Software Engineering}{8 - 13 November, 2020}{Sacramento, California, United States}
 \usepackage{framed}
\definecolor{shadecolor}{rgb}{0.9,0.9,0.9}
\def\ra{\textrightarrow}
\usepackage{xhfill}
\begin{document}

\maketitle

\section{Introduction}

\begin{figure*}[th]
    \centering
    \captionsetup[subfigure]{labelformat=empty}
    \begin{subfigure}{.33\linewidth}
        \centering
        \includegraphics[width=.8\linewidth]{images/evolution-apples}
        \caption{\texttt{`natural foods'} (.956)~~\ra{}~~\texttt{`granny smith'} (.986)}
    \end{subfigure}
    \hfill
    \begin{subfigure}{.33\linewidth}
        \centering
        \includegraphics[width=.8\linewidth]{images/evolution-snow}
        \caption{\texttt{`skiing'} (.937)~~\ra{}~~\texttt{`snow'} (.982)}
    \end{subfigure}
    \hfill
    \begin{subfigure}{.33\linewidth}
        \centering
        \includegraphics[width=.8\linewidth]{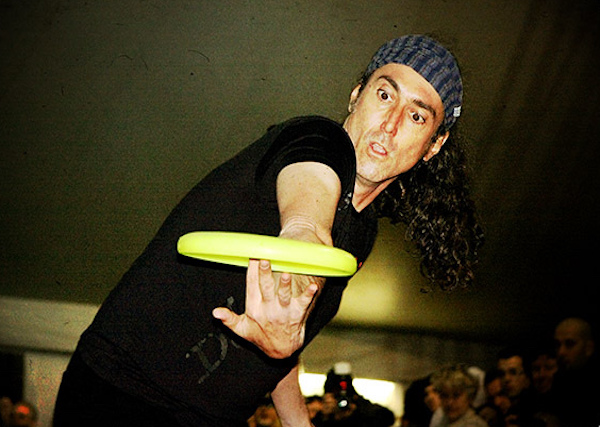}
        \caption{\texttt{`girl'} (.660)~~\ra{}~~\texttt{`photography'} (.738)}
    \end{subfigure}
    
    \bigskip
    
    \begin{subfigure}{.33\linewidth}
        \centering
        \includegraphics[width=.8\linewidth]{images/evolution-waves}
        \caption{\texttt{`water'} (.972)~~\ra{}~~\texttt{`wave'} (.932)}
    \end{subfigure}
    \hfill
    \begin{subfigure}{.33\linewidth}
        \centering
        \includegraphics[width=.8\linewidth]{images/evolution-tennis}
        \caption{\texttt{`tennis'} (.982)~~\ra{}~~\texttt{`sports'} (.989)}
    \end{subfigure}
    \hfill
    \begin{subfigure}{.33\linewidth}
        \centering
        \includegraphics[width=.8\linewidth]{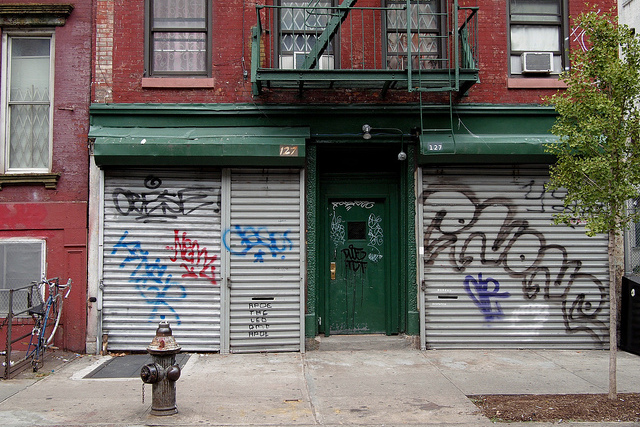}
        \caption{\texttt{`neighbourhood'} (.925)~~\ra{}~~\texttt{`blue'} (.927)}
    \end{subfigure}
    \caption{Prominent computer vision services evolve with time which is not effectively communicated to developers. Each image was uploaded in November 2018 and March 2019 and the topmost label was captured. Specialisation in labels (\textit{Left}), generalisation in labels (\textit{Centre}) and emphasis change in labels (\textit{Right}) are all demonstrated from the same service with no API change and limited release note documentation. Confidence values indicated in parentheses.}
    \label{fig:labelchanges}
\end{figure*}

The introduction of `intelligent' services into the software engineering ecosystem allows developers to leverage the power of AI without implementing complex ML algorithms, source and label training data, or orchestrate powerful and large-scale hardware infrastructure. This is extremely enticing for developers to embrace due to the effort, cost and non-trivial expertise required to implement AI in practice~\citep{Sculley2015, Polyzotis2018a}. 

However, the vendors that offer these services also periodically update their behaviour (responses). The ideal practice for communicating the evolution of a web service involves updating the version number and writing release notes. The release notes typically describe new capabilities, known problems, and requirements for proper operation \citep{SWEBOK}. Developers anticipate changes in behaviour between versioned releases although they expect the behaviour of a specific version to remain stable over time \citep{vasa2010growth}. However, emerging evidence indicates that `intelligent' services \textit{do not} communicate changes explicitly \citep{Cummaudo:2019esem}. Intelligent services evolve in unpredictable ways, provide no notification to developers and changes are undocumented~\citep{Cummaudo:2020icse}. To illustrate this, consider \cref{fig:labelchanges}, which shows the evolution of a popular computer vision service with examples of labels and associated confidence scores with how they changed. This behaviour change severely negatively affects reliability. Applications may no longer function correctly if labels are removed or confidence scores change beyond predefined thresholds. 

Unlike traditional web services, the functionality of these \textit{intelligent services} is dependent on a set of assumptions unique to their machine learning principles and algorithms. These assumptions are based on the data used to train machine learning algorithms, the choice of algorithm, and the choice of data processing steps---most of which are not documented to service end users. The behaviour of these services evolve over time~\citep{Cummaudo:2019icsme}---typically this  implies the underlying model has been updated or re-trained. 

Vendors do not provide any guidance on how best to deal with this evolution in client applications. For developers to discover the impact on their applications they need to know the behavioural deviation and the associated impact on the robustness and reliability of their system. Currently, there is no guidance on how to deal with this evolution, nor do developers have an explicit checklist of the likely errors and changes that they must test for~\citep{Cummaudo:2020icse}.

In this paper, we present a reference architecture to detect the evolution of such intelligent web services, using a mature subset of these services that provide computer vision as an exemplar. This tactic can be used both by intelligent service consumers, to defend their applications against the evolutionary issues present in intelligent web services, and by service vendors to make their services more robust. We also present a set of error conditions that occur in existing computer vision services.  

The key contributions of this paper are: 

\begin{itemize}
    \item A set of new service error codes for describing the empirically observed error conditions in intelligent services.
    \item A new reference architecture for using intelligent services with a Proxy Server that returns error codes based on an application specific benchmark dataset.
    \item A labelled data set of evolutionary patterns in computer vision services.
    \item An evaluation of the new architecture and tactic showing its efficacy for supporting intelligent web service evolution from both provider and consumer perspectives.
\end{itemize}

The rest of this paper is organised thus: \cref{fse2020:sec:motivation} presents a motivating example that anchors our work; \cref{sec:background} presents a landscape analysis on intelligent web services; \cref{fse2020:sec:solution} presents an overview of our architecture; \cref{fse2020:sec:eval} describes the technical evaluation; \cref{fse2020:sec:discussion} presents a discussion into the implications of our architecture, its limitations and potential future work; \cref{fse2020:sec:related-work} discusses related work; \cref{fse2020:sec:conclusions} provides concluding remarks.

\section{Motivating Example}
\label{fse2020:sec:motivation}

We identify the key requirements for managing evolution of intelligent services using a motivating example. Consider Pam, a software engineer tasked with developing a fall detector system for helping aged care facilities respond to falls promptly. Pam decides to build the fall detector with an intelligent service for detecting people as she has no prior experience with machine learning. The initial system built by Pam consists of a person detector and custom logic to identify a fall based on rapid shape deformation (i.e., a vertical `person' changing to a horizontal `person' greater than specified probability threshold value). Due to the inherent uncertainty present in an intelligent service and the importance of correctly identifying falls, Pam informs the aged care facility that they should manually verify falls before dispatching a nurse to the location. The aged care facility is happy with this approach but inform Pam that only a certain percentage of falls can be manually verified based on the availability of staff. In order to reduce the manual work Pam sets thresholds for a range of confidence scores where the system is uncertain. Pam completes the fall detector using a well-known cloud-based intelligent image classification web service and her client deploys this new fall detection application. 

Three months go by and then the aged care facility contact Pam saying the percentage of manual inspections is far too high and could she fix it. Pam is mystified why this is occurring as she thoroughly tested the application with a large dataset provided by the aged care facility. On further inspection Pam notices that the problem is caused by some images classifying the person with a \texttt{`child'} label rather than a \texttt{`person'} label. Pam is frustrated and annoyed at this behaviour as (i) the cloud vendor did not document or notify her of the change of the intelligent service behaviour, (ii) she does not know the best practice for dealing with such a service evolution, and (iii) she cannot predict how the service will change in the future. This experience also makes Pam wonder what other types of evolution can occur and how can she minimise these behavioural changes on her critical care application. Pam then begins building an ad-hoc solution hoping that what she designs will be sufficient. 

 For Pam to build a robust solution she needs to support the following requirements:

\begin{itemize}
    \item [\textbf{R1.}] Define a set of error conditions that specify the types of evolution that occur for an intelligent service.
    \item [\textbf{R2.}] Provide a notification mechanism for informing client applications of behavioural changes to ensure the robustness and reliability of the application. 
    \item [\textbf{R3.}] Monitor the evolution of intelligent services for changes that affect the application's behaviour. 
    \item [\textbf{R4.}] Implement a flexible architecture that is adaptable to different intelligent services and application contexts to facilitate reuse. 
\end{itemize}

\section{Intelligent Services}
\label{sec:background}

We present background information on intelligent services describing how they differ from traditional web services, the dimensions of their evolution and the currently limited configuration options available to users. 

\subsection{`Intelligent' vs `Traditional' Web Services}

Unlike conventional web services, intelligent web services are built using AI-based components. These components are unlike traditional software engineering paradigms as they are data-dependent and do not result in deterministic outcomes. These services make future predictions on new data based solely against its training dataset; outcomes are expressed as probabilities that the inference made matches a label(s) within its training data. Further, these services are often marketed as forever evolving and `improving'. This means that their large training datasets may continuously update the prediction classifiers making the inferences, resulting both in probabilistic and non-deterministic outcomes~\citep{Cummaudo:2019icsme,Hosseini:2018jr}. Critically for software engineers using the services, these non-deterministic aspects have not been sufficiently documented in the service's API documented, which has been shown to confuse developers~\citep{Cummaudo:2020icse}.

A strategy to combat such service changes, which we often observe in traditional software engineering practices, are for such services to be versioned upon substantial change. Unfortunately emerging evidence indicates that prominent cloud vendors providing these intelligent services do not release new versioned endpoints of the APIs when the \textit{internal model} changes~\citep{Cummaudo:2019icsme}. For intelligent services, it is impossible to invoke requests specific to a particular version model that was trained at a particular date in time.  This means that developers need to consider how evolutionary changes to the intelligent web services they make use of may impact their solutions \textit{in production}.

\subsection{Dimensions of Evolution}

\begin{figure}
    \centering
    \includegraphics[width=\linewidth]{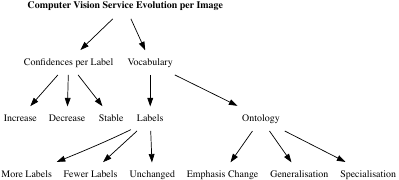}
    \caption{The dimensions of evolution identified within computer vision services.}
    \label{fig:service-evolution}
\end{figure}

The various key dimensions of the evolution of intelligent services is illustrated in \cref{fig:service-evolution}. There are two primary dimensions of evolution: \textit{changes to the label sets} returned per image submitted and \textit{changes to the confidences} per label in the set of labels returned per image. In the former, we identify two key aspects: cardinality changes and ontology changes. Cardinality changes occur when the service either introduces or drops a label for the same image at two different generations. Alternatively, the cardinality may remain stagnant, although this is not guaranteed. This results in an expectation mismatch by developers as to what labels can or will be returned by the service. For instance, the terms `black' and `black and white' may be found to be categorised as two separate labels. Secondly, the ontologies of these labels are non-static, and a label may become more generalised into a hypernym, specialised into a hyponym, or the emphasis of the label may change either to a co-hyponym or another aspect in the image, such as the colour or scene, rather than the subject of the image \citep{Cummaudo:2019icsme}.

\begin{figure}
    \centering
    \includegraphics[width=0.6\linewidth]{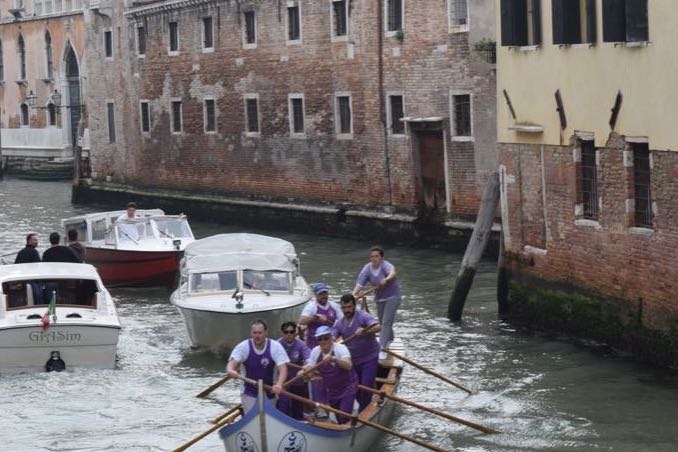}
    \caption{A significant confidence increase ($\delta = +0.425$) from \texttt{`window'} (0.559) to \texttt{`water transportation'} (0.984) goes beyond simple decision boundaries.}
    \label{fig:large-delta}
\end{figure}

Secondly, we have identified that the confidence values returned per label are also non-static. While some services may present minor changes to labels' confidences resulting from statistical noise, other labels had significant changes that were beyond basic decision boundaries. An example is shown in \cref{fig:large-delta}. Developer code written to assume certain ranges/confidence intervals will fail if the service evolves in this way.

\subsection{Limited Configurability}

\begin{figure}[t]
    \includegraphics[width=\linewidth]{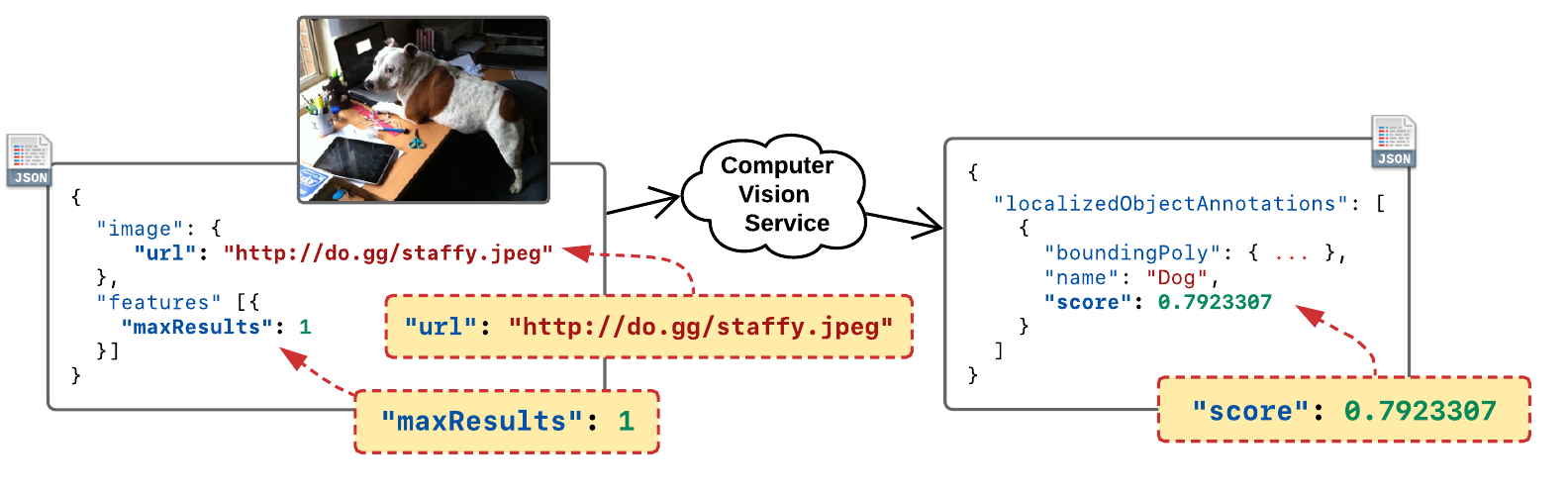}
    \caption{Request and response for an intelligent computer vision web service with only three configuration parameters: the image's \texttt{url}, \texttt{maxResults} and \texttt{score}.}
    \label{fig:dog-example}
\end{figure}
\begin{figure*}[t]
    \centering
    \includegraphics[width=.9\linewidth]{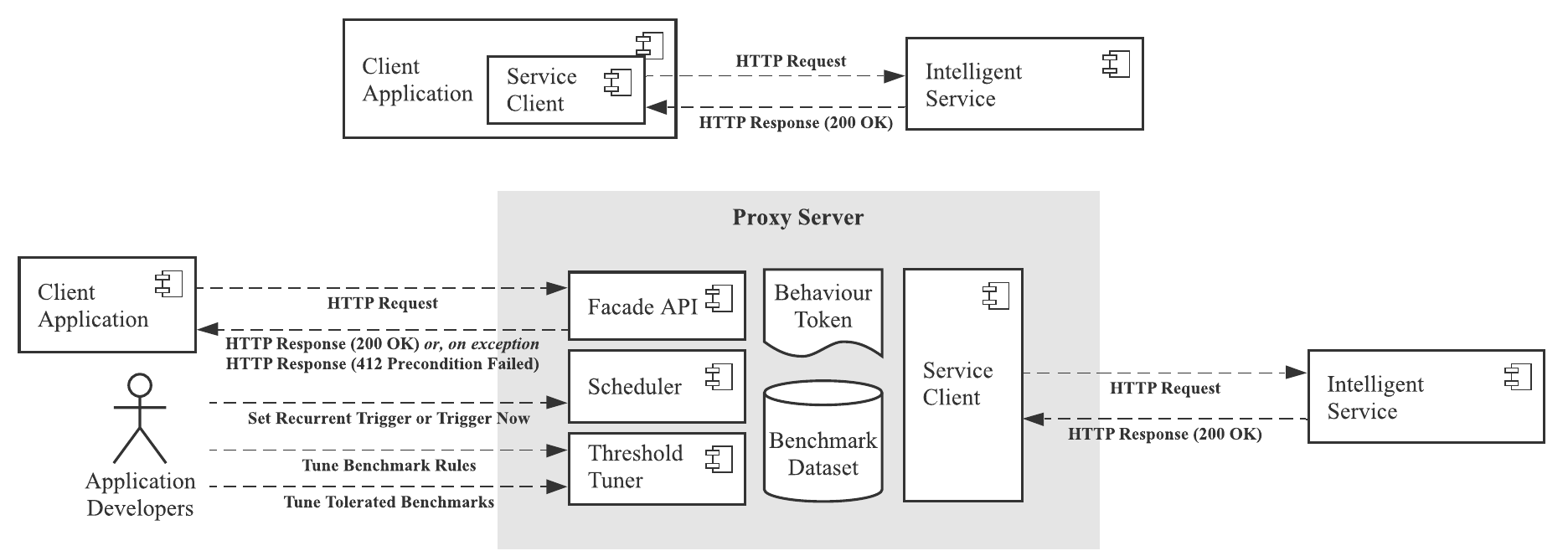}
    \caption{\textit{Top:} Accessing an intelligent service directly. \textit{Bottom:} Primary components of the Proxy Server approach.}
    \label{fig:facade-overview}
\end{figure*}
As an example, consider \cref{fig:dog-example}, which illustrates an image of a dog uploaded to a well-known cloud-based computer vision service. Developers have very few configuration parameters in the upload payload (\texttt{url} for the image to analyse and \texttt{maxResults} for the number of objects to detect). The JSON output payload provides the confidence value of its estimated bounding box and label of the dog object via its \texttt{score} field (0.792). This value indicates the level of confidence in the label returned, and is dependent on the input to the underlying ML model used by that service. Developers set thresholds as a decision boundary in this case, a threshold of ``greater than 0.7'' could indicate that the image contains a dog where as any other value the system is uncertain. These decision boundaries determine if the service's output is accepted or rejected. However, these confidence scores change whenever a model is re-trained and these changes are not communicated or propagated to developers \citep{Cummaudo:2019icsme}. Developers can only modify these decision boundaries to improve the performance of the intelligent web service. This is unlike many machine learning toolkit hyper-parameter optimisation facilities, which can be used to configure the internal parameters of the algorithm for training a model. In this case, developers using the intelligent web service have no insight into which hyperparameters were used when training the model or the algorithm selected, and cannot tune the trained model. Thus an evaluation procedure must be followed as a part of using an intelligent service for an application to tune their output confidence values and select appropriate threshold boundaries. While some service providers provide some guidance to thresholding,\footnoteurl{https://bit.ly/36oMgWb}{20 May 2020} they do not provide domain-specific tooling. This is because choice of appropriate thresholds is dependent on the data and must consider factors, such as algorithmic performance, financial cost, and impact of false-positives/negatives.

However, decision boundaries in service client code using simple \texttt{If} conditions around confidence scores is not a sufficient strategy, as evidence shows intelligent, non-deterministic web services change sporadically and unknowingly. Most traditional, deterministic code bases handle unexpected behaviour of called APIs via \textit{error codes} and exception handling. Thus the non-deterministic components of the client code, such as those using computer vision services, will also tend to conflict with their traditional deterministic components as the latter do not deal in terms of probabilities but in using error codes. 
This makes achieving robust component integration in client code bases hard.
More sophisticated monitoring of intelligent services in client code is therefore required to map the non-deterministic service behaviour changes to errors such that the surrounding infrastructure can support it and reduce interface boundary problems. While data science literature acknowledges the need for such an architecture \citep{10.1007/978-3-319-08976-8_16} they do not offer any technical software engineering solutions to mitigate the issues such that software engineers have a pattern to work against it. To date, there do not yet exist intelligent web service client code architectures, tactics or patterns that achieve this goal.

\begin{table*}[t!]
    \centering
    \caption{Potential reasons for a \texttt{412 Precondition Failed} response.}
    \label{tab:error-codes}
    \small
    \begin{tabular}{p{0.2\linewidth}|p{0.7\linewidth}}
        \toprule    
        \textbf{Error Code} &
        \textbf{Error Description} \\
        \midrule
        No Key Yet & This indicates that the Proxy Server is still initialising its first behaviour token, i.e., $k_{0}$ does not yet exist. \\
        Service Mismatch & The service encoded within the behaviour token provided to the Proxy Server does not match the service the Proxy Server is benchmarked against. This makes it possible for one Proxy Server to face multiple computer vision services. \\
        Dataset Mismatch & The benchmark dataset $B$ encoded within the behaviour token does not match the benchmark dataset encoded within the Proxy Server.  \\
        Success Mismatch & The success of each response within the benchmark dataset must be true for a behaviour token to be used within a request. This error indicates that $k_{r}$ is, therefore, not successful. \\
        Min Confidence Mismatch & The minimum confidence delta threshold set in $k_{t}$ does not match that of $k_{r}$. \\
        Max Labels Mismatch & The maximum label delta threshold set in $k_{t}$ does not match that of $k_{r}$. \\
        Response Length Mismatch & The number of responses within $k_{t}$ does not match that within $k_{r}$. \\
        \midrule
        Label Delta Mismatch & An image within $B$ has either dropped or gained a number of labels that exceeds the maximum label delta. Thus, $k_{r}$ exceeds the threshold encoded within $k_{t}$. \\
        Confidence Delta Mismatch & One of the labels within an image encoded in $k_{r}$ exceeds the confidence threshold encoded within $k_{t}$. \\
        Expected Labels Mismatch & One of the expected labels for an image within $k_{t}$ is now missing.\\
        \bottomrule
    \end{tabular}
\end{table*}

\section{Our Approach}
\label{fse2020:sec:solution}

To address the requirements from \cref{fse2020:sec:motivation} we have developed a new Proxy Service\footnote{A reference architecture is provided at \urlrefarch{}.} that includes: (i) evaluation of an intelligent service using an application specific benchmark dataset, (ii) a Proxy Server to provide client applications with evolution aware errors, and (iii) a scheduled evolution detection mechanism. The current approach of using an intelligent API via direct access is shown in \cref{fig:facade-overview} (top). In contrast, an overview of our approach is shown in \cref{fig:facade-overview} (bottom). The following sections describe our approach in detail. 

\subsection{Core Components}

For the purposes of this paper we assume that the intelligent service of interest is an image recognition service, but our approach generalises to other intelligent, trained model-based services e.g. NLP, document recognition, voice, etc. Each image, when uploaded to the intelligent service returns a response ($R$) which is a set describing a label ($l$) of what is in the image ($i$) along with its associated confidence ($c$)---thus $R_{i} = \{ (l_{1}, c_{1}), (l_{2}, c_{2}), \dots (l_{n}, c_{n}) \}$. Most documentation of these services imply that these confidence values are all what is needed to handle evolution in their systems. This means that if a label changes beyond a certain threshold, then the developer can deal with the issue then (or ignore it). While this approach may work in some simple application contexts, in many it may not. Our Proxy Server offers a way to monitor if these changes go beyond a threshold of tolerance, checking against a domain-specific dataset over time.

\subsubsection{Benchmark Dataset} Monitoring an intelligent service for behaviour change requires a Benchmark Dataset, a set of $n$ images. For each image ($i$) in the Benchmark Dataset ($B$) there is an associated label ($l$) that represents the true value for that item; $B_{i} = \{(i_{1}, l_{1}), (i_{2}, l_{2}), \dots (i_{n}, l_{n})\}$. This dataset is used to check for evolution in intelligent services by periodically sending each image within the dataset to the service's API, as per the rules encoded within the Scheduler (see \cref{fse2020:sec:components:scheduler}). By using a dataset specific to the application domain, developers can detect when evolution affects their application rather than triggering all non-impactful changes. This helps achieve our requirement \textit{R3. Monitor the evolution of intelligent services for changes that affect the application’s behaviour}. Using application-specific datasets also ensures that the architectural style can be used for different intelligent services as only the data used needs to change. This design choice encourages reuse, satisfying requirement \textit{R4. Implement a flexible architecture that is adaptable to different intelligent services and application contexts to facilitate reuse}. We propose an initial set of guidelines on how to create and update the benchmark dataset within \cref{fse2020:sec:future-work:guidelines}.

\subsubsection{Facade API} An architectural `facade' is the central component to our mitigation strategy for monitoring and detecting for changes in called intelligent services. The facade acts as a guarded gateway to the intelligent service that defends against two key issues: (i)~potential shifts in model variations that power the cloud vendor services, and (ii)~ensures that a context-specific dataset specific to the application being developed is validated \textit{over time}. By using a facade we can return evolution-aware error codes to the client application satisfying requirement \textit{R1. Define a set of error conditions that specify the types of evolution that occur for an intelligent service} and enabling requirement \textit{R3.  Monitor the evolution of intelligent services for changes that affect the application’s behaviour.} This works by ensuring every request made by the client application contains a valid Behaviour Token (see \cref{fse2020:sec:components:behaviour-token}) and will reject the request when evolution has been identified by the Scheduler with an associated error code. The Facade API essentially `blocks' the client application out from accessing the intelligent service when an invalid state has occurred.

\subsubsection{Threshold Tuner} Selecting an appropriate threshold for detecting behavioural change depends on the application context. Setting the threshold too low increases the likelihood of incorrect results, while setting the threshold too high means undesired changes are being detected. Our approach enables developers to configure these parameters through a Threshold Tuner. This improves robustness as now there is a systematic approach for monitoring and responding to incorrect thresholds. Configurable thresholds meet our key requirements \textit{R2} and \textit{R3}.  

\begin{table}[t]
    \centering
    \caption{Rules encoded within a Behaviour Token.}
    \label{tab:behaviour-token-rules}
    \small
    \begin{tabular}{p{0.25\linewidth}|p{0.675\linewidth}}
    \toprule
    \textbf{Rule} &
    \textbf{Description}\\
    \midrule
    Max Labels & The value of~$n$.\\
    Min Confidence & The smallest acceptable value of~$c$.\\
    Max $\delta$ Labels & The minimum number of labels dropped or introduced from the current $k_{t}$ and provided $k_{r}$ to be considered a violation (i.e $|l(k_{t})~\triangle~l(k_{r})|$).\\
    Max $\delta$ Confidence & The minimum confidence change of \textit{any} label from the current $k_{t}$ and provided $k_{r}$ to be considered a violation.\\
    Expected Labels & A set of labels that every response must include.\\
    \bottomrule
    \end{tabular}
\end{table}
\begin{figure*}[t]
    \centering
    \includegraphics[width=\linewidth]{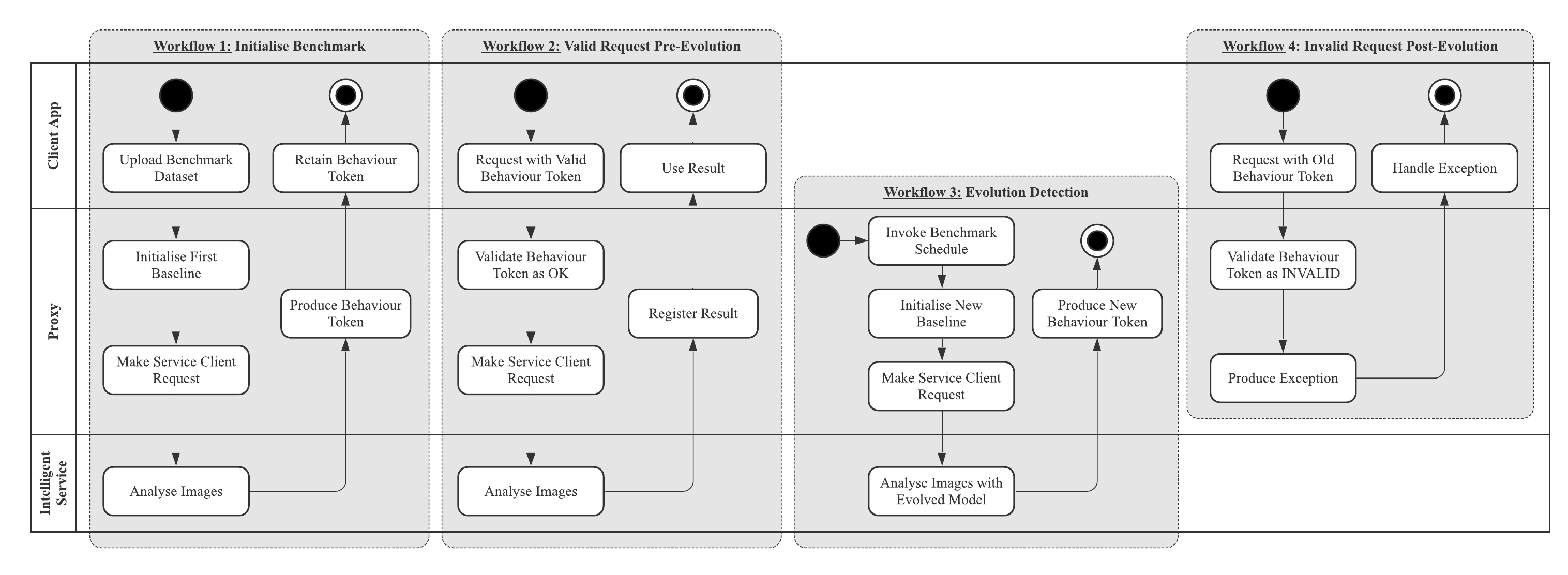}
    \caption{State diagram for the four workflows presented.}
    \label{fig:dynamic-behaviour}
\end{figure*}
\subsubsection{Behaviour Token} \label{fse2020:sec:components:behaviour-token}
The Behaviour Token stores the current state of the Proxy Server by encoding specific rules regarding the evolution of the intelligent service. The current token (at time $t$) held by the Proxy Server is denoted by $k_{t}$. These rules are specified by the developer upon initialisation of this Proxy Server, and are presented in \cref{tab:behaviour-token-rules}. When the Proxy Server is first initialised (i.e., at $t = 0$), the first Behaviour Token is created based on the Benchmark Dataset and its configuration parameters (\cref{tab:behaviour-token-rules}) and is stored locally (thus $k_{0}$ is created). The Behaviour Token is passed to the client application to be used in subsequent requests to the proxy server, where $k_{r}$ represents the Behaviour Token passed from the client application to the proxy server. Each time the proxy server receives the Behaviour Token from the client the validity of the token is validated with a comparison to the Proxy Server's current behaviour token (i.e., $k_{r} \equiv k_{t}$). An invalid token (i.e., when $k_{r} \not\equiv k_{t}$)  indicates that an error caused by evolution has occurred and the application developer needs to appropriately handle the exception. Behaviour Tokens are essential for meeting requirement \textit{R3. Monitor the evolution of intelligent services for changes that affect the application’s behaviour.}

\subsubsection{Service Client}

If any of the rules above are violated, then the response of the facade request will vary depending on the parameter of the behaviour encoded within the behaviour token. This can be one of:

\begin{itemize}
    \item \textbf{Error:} Where a HTTP non-200 code is returned by the facade to the client application, indicating that the client application must deal with the issue immediately;
    \item \textbf{Warning:} Where a warning `callback' endpoint is called with the violated response to be dealt with, but the response is still returned to the client application;
    \item \textbf{Info:} Where the violated response is logged in the facade's logger for the developer to periodically read and inspect, and the response is returned to the client application.
\end{itemize}

We implement this Proxy Server pattern using HTTP conditional requests. As we treat the Label as a first class citizen, we return the labels for a specific image ($r_{i}$) only where the \textit{Entity Tag} (ETag) or \textit{Last Modified} validators pass. The $k_{r}$ is encoded within either the ETag (i.e., a unique identifier representing $t$) or as the date labels (and thus models) were last modified (i.e., using the \texttt{If-Match} or \texttt{If-Unmodified-Since} conditional headers). We note that the use of \textit{weak} ETags should be used, as byte-for-byte equivalence is not checked but only semantic equivalence within the tolerances specified. Should $t$ evolve to an invalid state (i.e., $k_{r}$ is no longer valid against $k_{t}$) then the behaviour as described above will be enacted.

These HTTP header fields are used as the `backbone' to help enforce robustness of the services against evolutionary changes and context within the problem domain dataset. Responses from the service are forwarded to the clients when such rules are met, otherwise alternative behaviour occurs. For example, the most severe of violated erroneous behaviour is the `Error' behaviour. To enforce this rule, we advocate for use of the \texttt{412 Precondition Failed} HTTP error if a violation occurs, as a \texttt{If-*} conditional header was violated. An example of this architectural pattern with the `Error' behaviour is illustrated in \cref{fig:dynamic-behaviour}. 

\begin{figure*}[t]
    \centering
    \includegraphics[width=.9\linewidth]{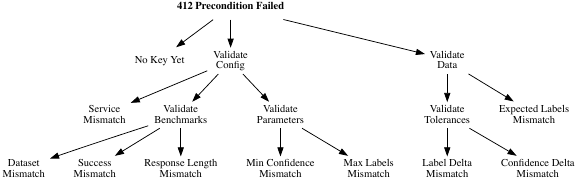}
    \caption{Precondition failure taxonomy; leaf nodes indicate error types returned to users.}
    \label{fig:precondition-failed}
\end{figure*}

We suggest the \texttt{412 Precondition Failed} HTTP error be returned in the event that a behaviour token is violated against a new benchmark. Further details outlining the reasons why a precondition has failed are encoded within a JSON response sent back to the consuming application. The following describes the two broad categories of possible errors returned: \textit{robustness precondition failure} or \textit{benchmark precondition failure}. These are illustrated in a high level within \cref{fig:precondition-failed} where leaf nodes are the potential error types that can be returned. A list of the different error codes are given in \cref{tab:error-codes}, where errors above the rule are robustness expectations (which check for basic requirements such as whether the key provided encodes the same data as the dataset in the facade) while those below are benchmark expectations (which identifies evolution cases).

\subsubsection{Scheduler} \label{fse2020:sec:components:scheduler} The Scheduler is responsible for triggering the Evolution Detection Workflow (described in detail below in \cref{fse2020:sec:workflow}). Developers set the schedule to run in the background at regular intervals (e.g., via a cron-job) or to trigger if violations occur $z$ times. The Scheduler is the component that enables our architectural style to identify called intelligent service software evolution and to notify the client applications that such evolution has occurred. Client applications can then respond to this evolution in a timely manner rather than wait for the system to fail, as in our motivating example. The Scheduler is necessary to satisfy our requirements \textit{R2} and \textit{R3}.

\subsection{Usage Example}
\label{fse2020:sec:workflow}

We explain how developer Pam, from our motivating example, would use our proposed solution to satisfy the requirements described in \cref{fse2020:sec:motivation}. Each workflow is presented in \cref{fig:dynamic-behaviour}. Only \textit{Workflow 1 - Initialise Benchmark} is executed once, while the rest are cycled. The description below assumes Pam has implemented the Proxy.

\subsubsection{Workflow 1. Initialise Benchmark}

The first task that Pam has to do is to prepare and initialise the benchmark dataset within the Proxy Server. To prepare a representative dataset, Pam needs to follow well established guidelines such as those proposed by \citet{pyle1999data}. Pam also needs to manually assign labels to each image before uploading the dataset to the Proxy along with the thresholds to use for detecting behavioural change. The full set of parameters that Pam has to set are based on the rules shown in \cref{tab:behaviour-token-rules}. Pam cannot use the Proxy to notify her of evolution until a Benchmark Dataset has been provided. The Proxy then sends each image in the Benchmark Dataset to the intelligent service and stores the results. From these results, a Behaviour Token is generated which is passed back to the Client Application. Pam uses this token in all future requests to the Proxy as the token captures the current state of the intelligent service.

\subsubsection{Workflow 2. Valid Request Pre-Evolution}

Workflow 2 represents the steps followed when the intelligent service is behaving as expected. Pam makes a request to label an image to the Proxy using the token that she received when registering the Benchmark Dataset. The token is validated with the Proxy's current state token and then a request to label the image is made to the intelligent service if no errors have occurred. Results returned by the intelligent service are registered with the Proxy Server. Pam can be confident that the result returned by our service is in line with her expectations.

\subsubsection{Workflow 3. Evolution Detection}

Workflow 3 describes how the Proxy functions when behavioural change is present in the called intelligent service. Pam sets a schedule for once a day so that the Proxy's Scheduler triggers Workflow 3. First, each image in the Benchmark Dataset is sent to the intelligent service. Unlike, Workflow 1, we already have a Behaviour Token that represents the previous state of the intelligent service. In this case, the model behind the intelligent service has been updated and provides different results for the Benchmark Dataset. Second, the Proxy updates the internal Behaviour Token ready for the next request. At this stage Pam will be notified that the behaviour of the intelligent service has changed. 

\subsubsection{Workflow 4. Invalid Request Post-Evolution}

Workflow 4 provides Pam with an error message when evolution has been detected. Pam's client application makes a request to the Proxy Server with an old Behaviour Token. The Proxy Server then validates the client token which is invalid as the Behaviour Token has been updated. In this case, an exception is raised and an appropriate error message as discussed above is included in the response back to Pam's client application. Pam can code her application to handle each error class in appropriate ways for her domain.

\section{Evaluation}
\label{fse2020:sec:eval}

Our evaluation of our novel intelligent service Proxy Server approach uses a technical evaluation based on the results of an observational study. We used existing datasets from observational studies~\citep{Cummaudo:2019icsme,Lin:2014vma} to identify problematic evolution in computer vision labelling services. This technical evaluation is designed to show: (i) what the responses are with and without our architecture present (highlighting errors); (ii) the overall increased robustness using enhanced responses; and (iii) the technical soundness of the approach. Thus, we propose the following research question which we answer in \cref{fse2020:eval:results}: \textit{``Can the architecture identify evolutionary issues of computer vision services via error codes?''} Based on our findings we proposed and implemented the Proxy Server using a Ruby development framework which we have made available online for experimentation.\footnoteurl{\urlrefarch{}}{5 March 2020} Additional data was collected from the computer vision service and sent to the Proxy Server to evaluate how the service handles behavioural change. 

\subsection{Data Collection and Preparation}

To minimise reviewer bias, we do not identify the name of the service used, however this service was one of the most adopted cloud vendors used in enterprise applications in 2018~\citep{RightScaleInc:2018kJ}. The two existing datasets used~\citep{Cummaudo:2019icsme,Lin:2014vma} consisted of 6,680 images. 

We initialised the benchmark (workflow 1) in November 2018, and sent each image to the service every eight days and captured the JSON responses through the facade API (workflow 2) until March 2019. This resulted in 146,960 JSON responses from the target computer vision service. We then selected the first and last set of JSON responses (i.e., 13,360 responses) and independently identified 331 cases of evolution of the original 6,680 images.  This was achieved by analysing the JSON responses for each image taken in using an evaluation script.\footnoteurl{http://bit.ly/2G7saFJ}{2 March 2020}

\begin{figure}[t]
    \centering
    \includegraphics[width=\linewidth]{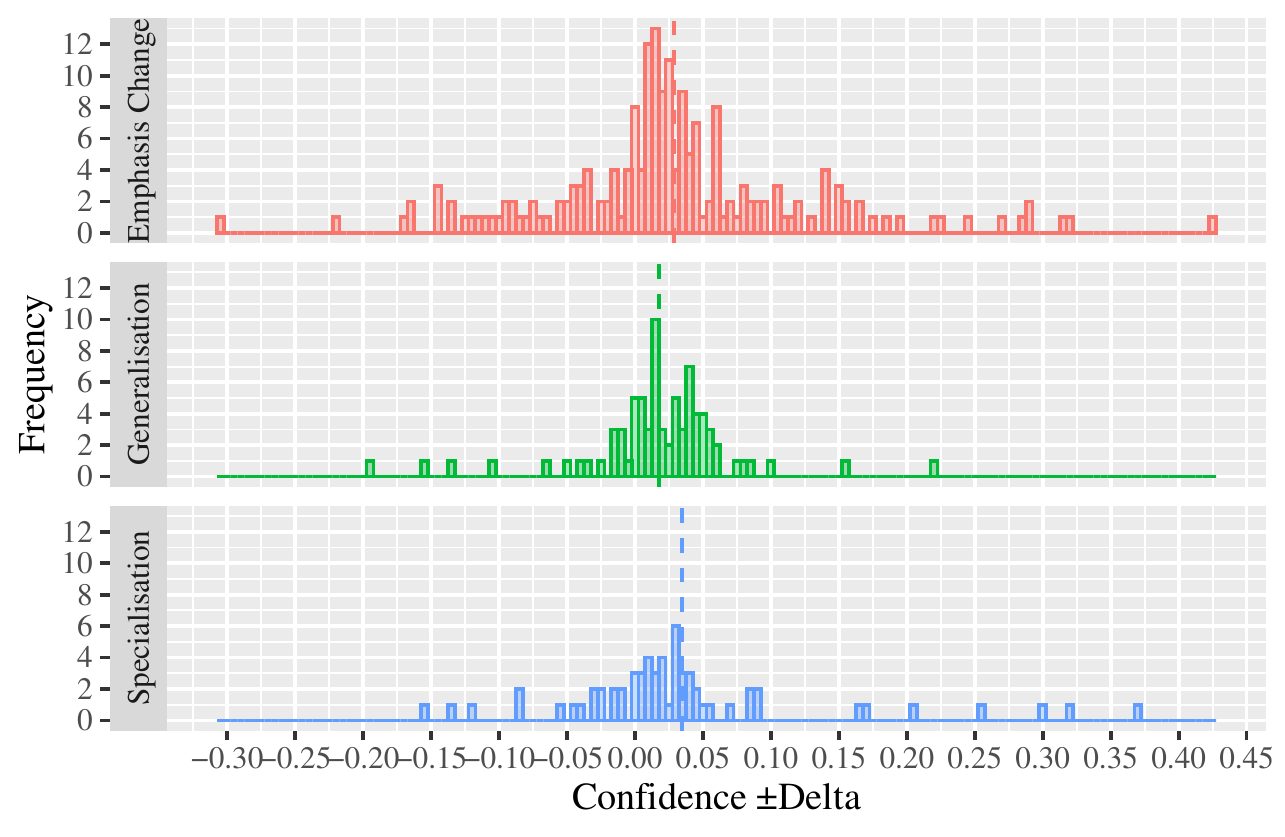}
    \caption{Histogram of confidence variation}
    \label{fig:frequency}
\end{figure}

For each JSON response, evolution (as classified by \cref{fig:service-evolution}) was determined either by a vocabulary or confidence per label change in the first and last responses sent. For the 331 evolving responses, we calculated the delta of the label's confidence between the two timestamps and the delta in the number of labels recorded in the entire response. Further, for the highest-ranking label (by confidence), we manually classified whether its ontology became more specific, more generalised or whether there was substantial emphasis change. The distribution of confidence differences per these three groups are shown in \cref{fig:frequency}, with the mean confidence delta indicated with a vertical dotted line. This highlights that, on average, labels that change emphasis generally have a greater variation, such as the example in \cref{fig:large-delta}. Further, we grouped each image into one of four broad categories---\textit{food}, \textit{animals}, \textit{vehicles}, \textit{humans}---and assessed the breakdown of ontology variance as provided in \cref{tab:ontology-variance}. We provide this dataset as an additional contribution and to permit replication.\footnoteurl{\urldataset{}}{5 March 2020}
The parameters set for our initial benchmark were a delta label value of 3 and delta confidence value of 0.01. Expected labels for relevant groups were also assigned as mandatory label sets (e.g., \textit{animal} images used \texttt{`animal'}, \texttt{`fauna'} and \texttt{`organism'}; \textit{human} images used \texttt{`human'} etc.).

\subsection{Results}
\label{fse2020:eval:results}

\begin{figure}
    \begin{framed}
    \centering
    \begin{minipage}{\linewidth}
        \begin{shaded*}
            \begin{minipage}{0.3\linewidth}
            \includegraphics[width=\linewidth]{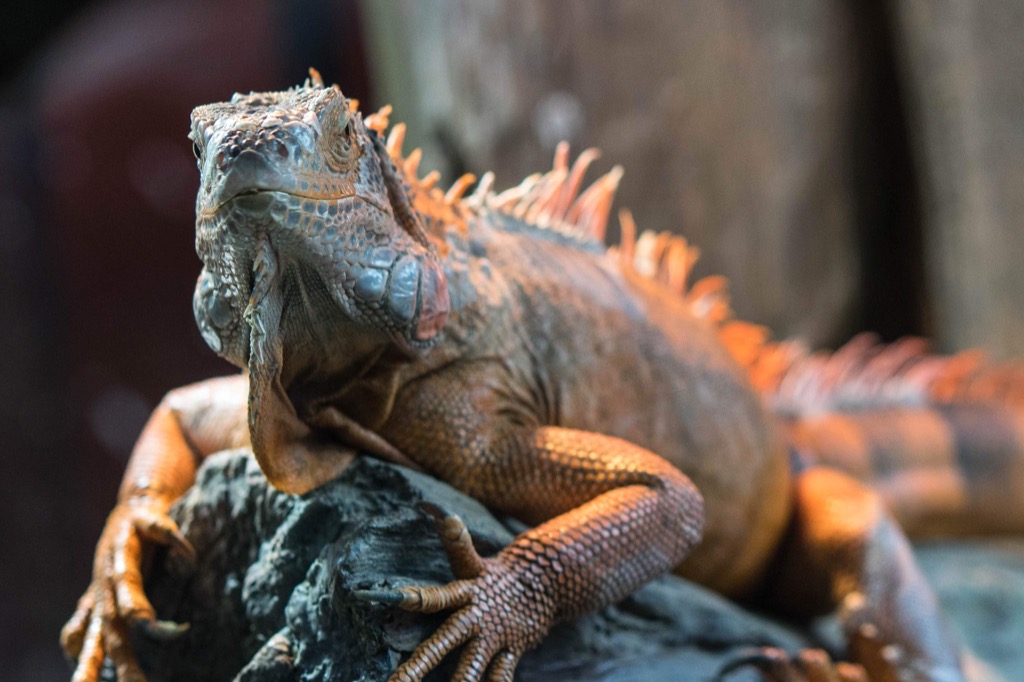}
            \end{minipage}
            \hfill
            \begin{minipage}{0.7\linewidth}
                \begin{tabular}{ll}
                \textbf{Label:}&Fauna\\
                \textbf{Nov 2018:}&\texttt{`reptile'} (0.9505)\\
                \textbf{Mar 2019:}&\texttt{`iguania'} (0.9836)\\
                \textbf{Category}:&Ontology Specialisation
                \end{tabular}
            \end{minipage}
        \end{shaded*}
    \end{minipage}
    \begin{minipage}{\linewidth}
        \centering\bigskip
        \noindent\xrfill[0.45ex]{.5pt}~~~\textit{Intelligent Service Response in March 2019}~~~\xrfill[0.45ex]{.5pt}
        \begin{lstlisting}[language=json]
{ "responses": [ { "label_annotations": [
  { "mid": "/m/08_jw6",
    "description": "iguania",
    "score": 0.9835183024406433,
    "topicality": 0.9835183024406433 },
  { "mid": "/m/06bt6",
    "description": "reptile",
    "score": 0.9833670854568481,
    "topicality": 0.9833670854568481 },
  { "mid": "/m/01vq7_",
    "description": "iguana",
    "score": 0.9796721339225769,
    "topicality": 0.9796721339225769 },
  ... ] } ] }
        \end{lstlisting}
    \end{minipage}
    \begin{minipage}{\linewidth}
        \noindent\xrfill[0.45ex]{.5pt}~~~\textit{Proxy Server Response in March 2019}~~~\xrfill[0.45ex]{.5pt}
        \begin{lstlisting}[language=json]
{ "error_code": 9,
  "error_type": "EXPECTED_LABELS_MISMATCH",
  "error_data": {
    "source_key": { ... },
    "violating_response": { ... },
    "uri": "http://localhost:4567/demo/data/0052.jpg",
    "expected_labels": [ "fauna" ],
    "labels_detected": [ "iguana", "green iguana", "iguanidae", "lizard", "scaled reptile", "marine iguana", "terrestrial animal", "organism" ],
    "labels_missing": [ "fauna" ],
    "reason": "The expected label(s) `fauna' are missing in the response." } }
        \end{lstlisting}
    \end{minipage}
    \end{framed}
    \caption{Example of an expected label missing due to evolution}
    \label{fig:example-expected-labels}
\end{figure}

Examples of the March 2019 responses contrasting the proxy and direct service responses in our evaluation are shown in \cref{fig:example-label-delta,fig:example-confidence-delta,fig:example-expected-labels}. (Due to space limitations, the entire JSON response is partially redacted using ellipses.) These examples identify the label identified with the highest level of confidence in three examples against the ground truth label in the benchmark dataset. In total, the Proxy Server identified 1,334 labels added to the responses and 1,127 labels dropped, with, on average, a delta of 8 labels added. The topmost labels added were \texttt{`architecture'} at 32 cases, \texttt{`building'} at 20 cases and \texttt{`ingredient'} at 20 cases; the topmost labels dropped were \texttt{`tree'} at 21 cases, \texttt{`sky'} at 19 cases and \texttt{`fun'} at 17 cases. 1054 confidence changes were also observed by the Proxy Server, on average a delta increase of 0.0977. 

In \cref{fig:example-confidence-delta}, we highlight an image of a sheep that was identified as a \texttt{`sheep'} (at 0.9622) in November 2018 and then a \texttt{`mammal'} in March 2019. This evolution was classified by the Proxy Server as a confidence change error as the delta in the confidences between the two timestamps exceeds the parameter set of 0.01---in this case, \texttt{`sheep'} was downgraded to the third-ranked label at 0.9816, thereby increasing by a value of 0.0194. As shown in the example, four other labels evolved for this image between the two time stamps (\texttt{`herd'}, \texttt{`livestock'}, \texttt{`terrestrial animal'} and \texttt{`snout'}) with an average increase of 0.1174 found. Such information is encoded as a 412 HTTP error returned back to the user by the Proxy Server, rejecting the request as substantial evolution has occurred, however the response \textit{directly} from the service indicates no error at all (indicating by a 200 HTTP response).

Similarly, \cref{fig:example-label-delta} shows a violation of the number of acceptable changes in the number of labels a response should have between two timestamps. In November 2018, the response includes the labels \texttt{`car'}, \texttt{`motor vehicle'}, \texttt{`city'} and \texttt{`road'}, however these labels are not present in the 2019 response. The response in 2019 introduces \texttt{`transport'}, \texttt{`building'}, \texttt{`architecture'}, and \texttt{`house'}. Therefore, the combined delta is 4 dropped and 4 introduced labels, exceeding our threshold set of 3.

Lastly, \cref{fig:example-expected-labels} indicates an expected label failure. In this example, the label \texttt{`fauna'} was dropped in the 2018 label set, which was an expected label of all animals we labelled in our dataset. Additionally, this particular response introduced  \texttt{`green iguana'}, \texttt{`iguanidae'}, and \texttt{`marine iguana'} to its label set. Therefore, not only was this response in violation of the label delta mismatch, it was also in violation of the expected labels mismatch error, and thus is caught twice by the Proxy Server.

\begin{figure}
    \begin{framed}
    \centering
    \begin{minipage}{\linewidth}
        \begin{shaded*}
            \begin{minipage}{0.3\linewidth}
            \includegraphics[width=\linewidth]{images/000000005992.jpg}
            \end{minipage}
            \hfill
            \begin{minipage}{0.7\linewidth}
                \begin{tabular}{ll}
                \textbf{Label:}&Animal\\
                \textbf{Nov 2018:}&\texttt{`sheep'} (0.9622)\\
                \textbf{Mar 2019:}&\texttt{`mammal'} (0.9890)\\
                \textbf{Category}:&Confidence Change
                \end{tabular}
            \end{minipage}
        \end{shaded*}
    \end{minipage}
    \begin{minipage}{\linewidth}
        \centering\bigskip
        \noindent\xrfill[0.45ex]{.5pt}~~~\textit{Intelligent Service Response in March 2019}~~~\xrfill[0.45ex]{.5pt}
        \begin{lstlisting}[language=json]
{ "responses": [ { "label_annotations": [
  { "mid": "/m/04rky",
    "description": "mammal",
    "score": 0.9890478253364563,
    "topicality": 0.9890478253364563 },
  { "mid": "/m/09686",
    "description": "vertebrate",
    "score": 0.9851104021072388,
    "topicality": 0.9851104021072388 },
  { "mid": "/m/07bgp",
    "description": "sheep",
    "score": 0.9815810322761536,
    "topicality": 0.9815810322761536 },
  ... ] } ] }
        \end{lstlisting}
    \end{minipage}
    \begin{minipage}{\linewidth}
        \noindent\xrfill[0.45ex]{.5pt}~~~\textit{Proxy Server Response in March 2019}~~~\xrfill[0.45ex]{.5pt}
        \begin{lstlisting}[language=json]
{ "error_code": 8,
  "error_type": "CONFIDENCE_DELTA_MISMATCH",
  "error_data": {
    "source_key": { ... },
    "source_response": { ... },
    "violating_key": { ... },
    "violating_response": { ... },
    "delta_confidence_threshold": 0.01,
    "delta_confidences_detected": {
      "sheep": 0.01936030388219212,
      "herd": 0.15035879611968994,
      "livestock": 0.13112884759902954,
      "terrestrial animal": 0.1791478991508484,
      "snout": 0.10682523250579834
    },
    "uri": "http://localhost:4567/demo/data/000000005992.jpeg",
    "reason": "Exceeded confidence delta threshold ±0.01 in 5 labels (delta mean=+0.1174)." } }
        \end{lstlisting}
    \end{minipage}
    \end{framed}
    \caption{Example of substantial confidence change due to evolution}
    \label{fig:example-confidence-delta}
\end{figure}
\begin{figure}
    \begin{framed}
    \centering
    \begin{minipage}{\linewidth}
        \begin{shaded*}
            \begin{minipage}{0.3\linewidth}
            \includegraphics[width=\linewidth]{images/000000019109.jpeg}
            \end{minipage}
            \hfill
            \begin{minipage}{0.7\linewidth}
                \begin{tabular}{ll}
                \textbf{Label:}&Vehicle\\
                \textbf{Nov 2018:}&\texttt{`vehicle'} (0.9045)\\
                \textbf{Mar 2019:}&\texttt{`motorcycle'} (0.9534)\\
                \textbf{Category}:&Label Set Change
                \end{tabular}
            \end{minipage}
        \end{shaded*}
    \end{minipage}
    \begin{minipage}{\linewidth}
        \centering\bigskip
        \noindent\xrfill[0.45ex]{.5pt}~~~\textit{Intelligent Service Response in March 2019}~~~\xrfill[0.45ex]{.5pt}
        \begin{lstlisting}[language=json]
{ "responses": [ { "label_annotations": [
  { "mid": "/m/07yv9",
    "description": "vehicle",
    "score": 0.9045347571372986,
    "topicality": 0.9045347571372986 },
  { "mid": "/m/07bsy",
    "description": "transport",
    "score": 0.9012271165847778,
    "topicality": 0.9012271165847778 },
  { "mid": "/m/0dx1j",
     "description": "town",
     "score": 0.8946694135665894,
     "topicality": 0.8946694135665894 },
  ... ] } ] }
        \end{lstlisting}
    \end{minipage}
    \begin{minipage}{\linewidth}
        \centering
        \noindent\xrfill[0.45ex]{.5pt}~~~\textit{Proxy Server Response in March 2019}~~~\xrfill[0.45ex]{.5pt}
        \begin{lstlisting}[language=json]
{ "error_code": 7,
  "error_type": "LABEL_DELTA_MISMATCH",
  "error_data": {
    "source_key": { ... },
    "source_response": { ... },
    "violating_key": { ... },
    "violating_response": { ... },
    "delta_labels_threshold": 5,
    "delta_labels_detected": 8,
    "uri": "http://localhost:4567/demo/data/000000019109.jpg",
    "new_labels": [ "transport", "building",   "architecture", "house" ],
    "dropped_labels": [ "car", "motor vehicle", "city",   "road" ],
    "reason": "Exceeded label count delta threshold ±5 (4 new labels + 4 dropped labels = 8)." } }
        \end{lstlisting}
    \end{minipage}
    \end{framed}
    \caption{Example of substantial changes of a response's label set due to evolution}
    \label{fig:example-label-delta}
\end{figure}

\begin{table}[]
    \centering
    \caption{Variance in ontologies for the five broad categories}
    \label{tab:ontology-variance}
    \tablefit{\begin{tabular}{c|ccccc|c}
        \toprule
        \textbf{Ontology Change}
        & \textbf{Food}
        & \textbf{Animal} 
        & \textbf{Vehicles} 
        & \textbf{Humans} 
        & \textbf{Other} 
        & \textbf{Total}\\
        \midrule
        Generalisation & 8&13&11&8&38&78\\
        Specialisation & 5&12&1&1&43&62\\
        Emphasis Change & 18&4&10&21&138&191\\
        \midrule
        Total & 31&29&22&30&219&331\\
        \bottomrule
    \end{tabular}}
\end{table}

\subsection{Threats to Validity}
\label{sec:threats}

\subsubsection*{Internal Validity} As mentioned, we selected a popular computer vision service provider to test our proxy server against. However, there exist many other computer vision services, and due to language barriers of the authors, no non-English speaking service were selected despite a large number available from Asia. Further, no user evaluation has been performed on the architectural tactic so far, and therefore developers may suggest improvements to the approach we have taken in designing our tactic. We intend to follow this up with a future study.

\subsubsection*{External Validity}
This paper only evaluates the object detection endpoint of a computer vision-based intelligent service. While this type of intelligent service is one of the more mature AI-based services available on the market---and is largely popular with developers \citep{Cummaudo:2020icse}---further evaluations of the our tactic may need to be explored against other endpoints (i.e., object localisation) or, indeed, other types of services, such as natural language processing, audio transcription, or on time-series data. Future studies may need to explore this avenue of research.

\subsubsection*{Construct Validity}
The evaluation of our experiment was largely conducted under clinical conditions, and a real-world case study of the design and implementation of our proposed tactic would be beneficial to learn about possible side-effects from implementing such a design (e.g., implications to cost etc.). Therefore, our evaluation does not consider more practical considerations that a real-world, production-grade system may need to consider.

\section{Discussion}
\label{fse2020:sec:discussion}

\subsection{Implications}

\subsubsection{For cloud vendors} Cloud vendors that provide intelligent services may wish to adopt the architectural tactic presented in this paper by providing a proxy, auxiliary service (or similar) to their existing services, thereby improving the current robustness of these services. Further, they should consider enabling developers of this technical domain knowledge by preventing client applications from using the service without providing a benchmark dataset, such that the service will return HTTP error codes.
These procedures should be well-documented within the service's API documentation, thereby indicating to developers how they can build more robust applications with their intelligent services.
Lastly, cloud vendors should consider updating the internal machine learning models less frequently unless substantial improvements are being made. Many different applications from many different domains are using these intelligent services so it is unlikely that the model changes are improving all applications. Versioned endpoints would help with this issue, although---as we have discussed---context using benchmark datasets should be provided.

\subsubsection{For application developers} Developers need to monitor all intelligent services for evolution using a benchmark dataset and application specific thresholds before diving straight into using them. It is clear that the evolutionary issues have significant impact in their client applications~\citep{Cummaudo:2019icsme}, and therefore they need to check the extent this evolution has  between versions of an intelligent service (should versioned APIs be available). Lastly, application developers should leverage the concept of a proxy server (or other form of intermediary) when using intelligent services to make their applications more robust.

\subsubsection{For project managers} Project managers need to consider the cost of evolution changes on their application when using intelligent services, and therefore should schedule tasks for building maintenance infrastructure to detect evolution. 
Consider scheduling tasks that evaluates and identifies the frequency of evolution for the specific intelligent service being used. Our research we have found some intelligent services that are not versioned but rarely show behavioural changes due to evolution.

\subsection{Limitations}

In the situation where a solo developer implements the Proxy Service the main limitation is the cost vs response time trade-off. Developers may want to be notified as soon as possible when a behavioural change occurs which requires frequent validation of the Benchmark Dataset. Each time the Benchmark Dataset is validated each item is sent as a request to the intelligent service. As cloud vendors charge per request to an intelligent service there are financial implications for operating the Proxy Service. If the developer optimises for cost then the application will take longer to respond to the behavioural change potentially impact end users. Developers need to consider the impact of cost vs response time when using the Proxy Service. 

Another limitation of our approach is the development effort required to implement the Proxy Service. Developers need to build a scheduling component, batch processing pipeline for the Benchmark Dataset, and a web service. These components require developing and testing which impact project schedules and have maintenance implications. Thus, we advise developers to consider the overhead of a Proxy Service and way up the benefits with have incorrect behaviour caused by evolution of intelligent services. 

\subsection{Future Work}

\subsubsection{Guidelines to construct and update the Benchmark Dataset}\label{fse2020:sec:future-work:guidelines} Our approach assumes that each category of evolution is present in the Benchmark Dataset prepared by the developer. Further guidelines are required to ensure that the developer knows how to validate the data before using the Proxy Service. While the focus of this paper was to present and validate our architectural tactic, guidelines on how to construct and update benchmark datasets for this tactic will need to be considered in future work. Data science literature extensively covers dataset preparation (e.g., \citep{pyle1999data,Krig2016}), and many example benchmark datasets are readily available \citep{wiki:dataset-list,guo2016ms,10.1145/2370216.2370437}. An initial set of guidelines are proposed as follows: data must be contextualised and appropriately sampled to be representative of the client application in particular the patterns present in the data, contain both positive and negative examples (this is/is not a cat); where to source data from (existing datasets, Google Images/Flickr, crowdsourced etc.); whether the dataset is synthetically generated to increase sample size; and how large a benchmark dataset size should be (i.e., larger the better but takes more effort and costs more). Benchmark datasets can also be used by software engineers provided the domain and context is appropriate for their specific application's context. Software engineers also benefit from our approach even if these guidelines are not strictly adhered to provided they use an application-specific dataset (i.e., data collected from the input source for their application). The main reason for this is that without our proposed tactic there are limited ways to build robust software with intelligent services. Future testing and evaluation of these guidelines should be considered.

\subsubsection{Extend the evolution categories to support other intelligent services} This paper has used computer vision services to assess our proposed tactic, and therefore further investigation is needed into the evolution characteristics of other intelligent services. The evolution challenges with services that provide optimisation algorithms such as route planning are likely to differ from computer vision services. These characteristics of an application domain have shown to greatly influence software architecture~\citep{barnett2018extracting} and further development of the Proxy Service will need to account for these differences. As an example, we have identified many similar issues that exist for natural language processing (NLP), where topic modelling produces labels on large bodies of text with associated confidences. Therefore, the \textit{broader} concepts of our contribution (e.g., behaviour token parameters, error codes etc.) can be used to handle issues in NLP to demonstrate the generalisability of the architecture to other intelligent services. We plan to apply our tactic to NLP and other intelligent services  in our future work.

\subsubsection{Provide tool support for optimising parameters for an application context} Appropriately using the Proxy Service requires careful selection of thresholds, benchmark rules and scheduling. Further work is required to support the developer in making these decisions so an optimal application specific outcome is achieved. One approach is to present the trade-offs to the developer and let them visualise the impact of their decisions. 

\subsubsection{Improvements for a more rigorous approach} Conducting a more formal evaluation of our proposed architecture would benefit the robustness of the solution presented. This could be done in various ways, for example, using a formal architecture evaluation method such as ATAM \citep{Kazman2000} or a similar variant \citep{Bouwers2010}; conducting user evaluation via brainstorming sessions or interviews with practitioners who may provide suggestions to improve our approach; determining better strategies to fully-automate the approach and reduce manual steps; and using real-world industry case studies to identify other factors such as cost and maintenance issues. All these are various avenues of research that would ultimately benefit in a more well-rounded approach to the architectural tactic we have proposed.

\section{Related Work}
\label{fse2020:sec:related-work}

\subsubsection*{Robustness of Intelligent Services}
While usage of intelligent services have been proven to have widespread benefits to the community~\citep{Reis:2018cp,daMotaSilveira:2017vp}, they are still largely understudied in software engineering literature, particularly around their robustness in production-grade systems. As an example, advancements in computer vision (largely due to the resurgence of convolutional neural networks in the late 1990s~\citep{Lecun:1998hy}) have been made available through intelligent services and are given marketed promises from prominent cloud vendors, e.g. ``with Amazon Rekognition, you don’t have to build, maintain or upgrade deep learning pipelines''.\footnote{\url{https://aws.amazon.com/rekognition/faqs/}, accessed 21 November 2019.} However, while vendors claim this, the state of the art of \textit{computer vision itself} is still susceptible to many robustness flaws, as highlighted by many recent studies~\citep{Eykholt:2018vk,Wang:2018vl,Rosenfeld:2018ut}. Further, each service has vastly different (and incompatible) ontologies which are non-static and evolve~\citep{Cummaudo:2019icsme,Ohtake:2019vi}, certain services can mislabel images when as little as 10\% noise is introduced~\citep{Hosseini:2018jr}, and developers have a shallow understanding of the fundamental ML concepts behind these issues, which presents a dichotomy of their understanding of the technical domain when contrasted to more conventional domains such as mobile application development~\citep{Cummaudo:2020icse}.

\subsubsection*{Proxy Servers as Fault Detectors}
Fault detection is an availability tactic that encompasses robustness of software~\citep{Bass:2003wi}. Our architecture implements the sanity check and condition monitoring techniques to detect faults~\citep{Bass:2003wi,Ingeno:2018}, by validating the reasonableness of the response from the intelligent service against the conditions set out in the rules encoded in the benchmark dataset and behaviour token. As we do in this study, the proxy server pattern can be used to both detect and action faults in another service as an intermediary between a client and a server. For example, addressing accessibility issues using proxy servers has been widely addressed~\citep{Zhang:2017,Takagi2000,Bigham2006,Bigham2008} and, more recently, they have been used to address in-browser JavaScript errors~\citep{Durieux2018}.

\section{Conclusions}
\label{fse2020:sec:conclusions}

Intelligent web services are gaining traction in the developer community, and this is shown with an evermore growing adoption of computer vision services in applications. These services make integration of AI-based components far more accessible to developers via simple RESTful APIs that developers are familiar with, and offer forever-`improving' object localisation and detection models at little cost or effort to developers. However, these services are dependent on their training datasets and do not return consistent and deterministic results. To enable robust composition, developers must deal with the evolving training datasets behind these components and consider how these non-deterministic components impact their deterministic systems.

This paper proposes an integration architectural tactic to deal with these issues by mapping the evolving and probabilistic nature of these services to deterministic error codes. We propose a new set of error codes that deal directly with the erroneous conditions that has been observed in intelligent services, such as computer vision. We provide a reference architecture via a proxy server that returns these errors when they are identified, and evaluate our architecture, demonstrating its efficacy for supporting intelligent web service evolution. Further, we provide a labelled dataset of the evolutionary patterns identified, which was used to evaluate our architecture.

\bibliography{references}

\end{document}